\documentclass[epj]{webofc}
\usepackage[varg]{txfonts}   
\usepackage{hyperref}
\usepackage{units}
\usepackage{siunitx}

\renewcommand*{\eqref}[1]{Eq.~(\ref{eq:#1})}

\newcommand*{\seclab}[1]{\label{sec:#1}}

\wocname{ARENA-2016}

\woctitle{ARENA-2016}

\begin{document}

\title{Results and Perspectives of the Auger Engineering Radio Array}

\author{\firstname{Christian} \lastname{Glaser}\inst{1}\fnsep\thanks{\email{glaser@physik.rwth-aachen.de}} for the Pierre Auger Collaboration\inst{2}}

\institute{RWTH Aachen University, III. Physikalisches Institut A, Aachen, Germany
\and  Full author list: \url{http://auger.org/archive/authors_2016_06.html}}

\abstract{
  The Auger Engineering Radio Array (AERA) is an extension of the Pierre Auger Cosmic-Ray Observatory. It is used to detect radio emission from extensive air showers with energies beyond \SI{E17}{eV} in the \unit[30 - 80]{MHz} frequency band. After three phases of deployment, AERA now consists of more than 150 autonomous radio stations with different spacings, covering an area of about \unit[17]{km$^2$}. It is located at the same site as other Auger low-energy detector extensions enabling combinations with various other measurement techniques. The radio array allows different technical schemes to be explored as well as cross-calibration of our measurements with the established baseline detectors of the Auger Observatory. We report on the most recent technological developments and give an overview of the experimental results obtained with AERA. In particular, we will present the measurement of the radiation energy, i.e., the amount of energy that is emitted by the air shower in the form of radio emission, and its dependence on the cosmic-ray energy by comparing with the measurement of the the well-calibrated Auger surface detector. Furthermore, we outline the relevance of this result for the absolute calibration of the energy scale of cosmic-ray observatories.
}

\maketitle

\section{The Auger Engineering Radio Array}
\seclab{sec-1}

The Auger Engineering Radio Array (AERA) is part of the Pierre Auger Observatory \cite{Auger2014} and measures extensive air showers with energies above \SI{E17}{eV}. 
The Pierre Auger Observatory is the world's largest detector for ultra-high energy cosmic rays. It is located in western Argentina and covers an area of \SI{3000}{km^2}. Its two baseline detectors are a surface detector (SD) array of more than 1600 water-Cherenkov detectors and 27 fluorescence telescopes (FD) that overlook that array from four different sites. In addition, underground scintillators are installed at the AERA site that allow for a separate measurement of the muonic air-shower component.

\begin{figure}[t]
\centering
\sidecaption
\includegraphics[width=0.75\textwidth,clip]{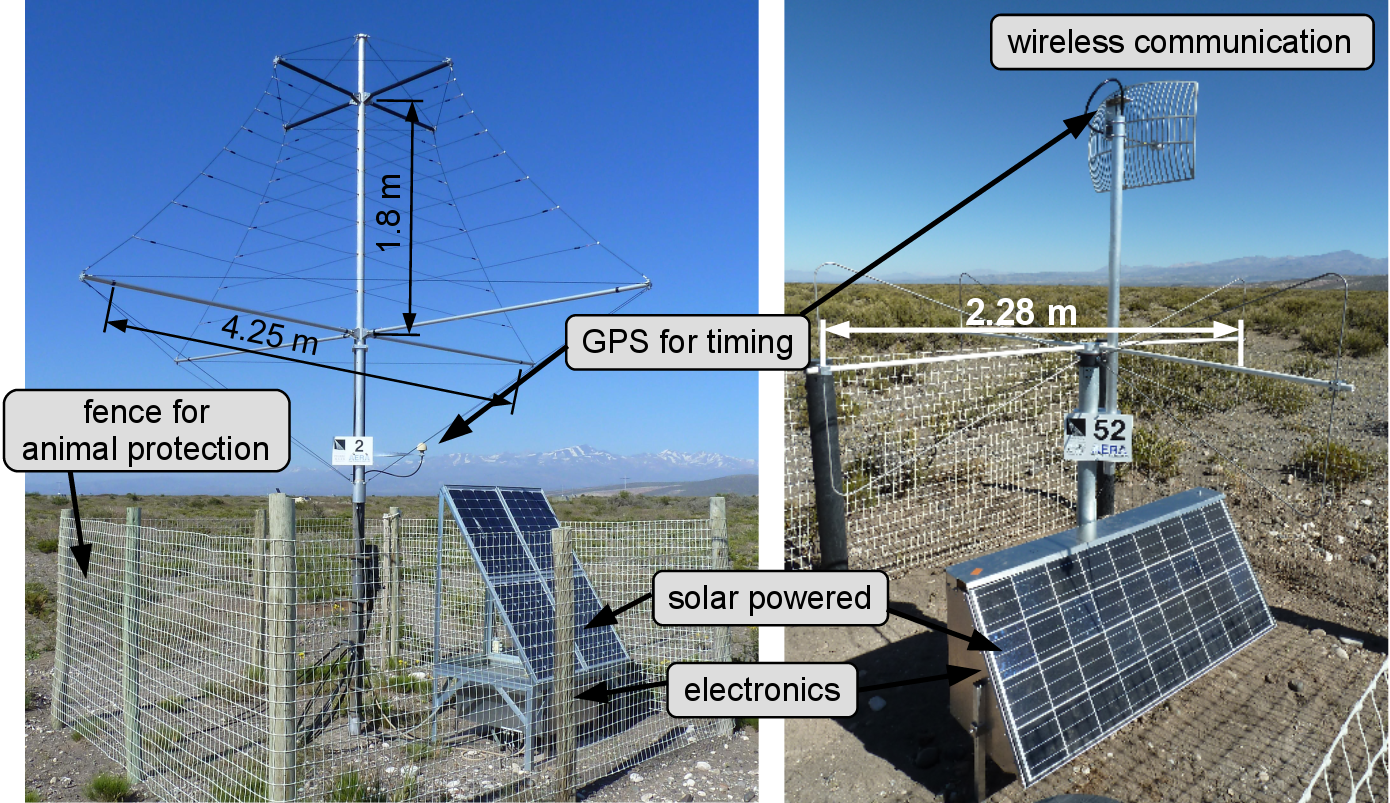}
\caption{Pictures of the two station designs deployed at AERA, left: log-periodic dipole antenna (LPDA), right: Butterfly antenna. }
\label{fig:antennas}
\end{figure}

In its current stage of expansion, AERA consists of 153 autonomously operating radio detector stations covering an area of \SI{17}{km^2}. Pictures of the two station designs deployed at AERA are shown in Fig.~\ref{fig:antennas}. Each station is solar powered and equipped with a battery to run \SI{24}{h} each day. A fence is put around each station to protect it from animals living in the Argentinian Pampa and the stations are equipped with GPS receivers for accurate timing. In addition, a method was develop to calibrate the relative timing of the radio stations to better than \unit[2]{ns} using a reference beacon and radio signals from commercial airplanes \cite{BeaconAirplane2016, ARENA2016Huege}. 

The first 24 stations that were installed in 2011 are equipped with log-periodic dipole antennas (LPDAs). The antenna response to cosmic-ray signals was previously simulated and measured extensively to an accuracy of 12.5\% \cite{AntennaPaper, Krause2016}. The stations are placed on a triangular grid with a distance of \SI{144}{m} covering an area of \SI{0.6}{km^2} and are connected via optical fibers to the data acquisition. 
All other AERA stations are equipped with a Butterfly antenna and were installed in 2013 (100 additional stations on a \SIlist{250;375}{m} grid) and 2015 (29 additional stations on a \SI{750}{m} grid). Instead of an optical fiber network the stations are equipped with a wireless communication system operating in the \SI{5}{GHz} band. 
All stations operate in the band from \unit[30 to 80]{MHz} and the radio signal is digitized with \SI{180}{MHz} or \SI{200}{MHz} depending on the type of digitizing hardware.

\begin{figure}[t]
\centering
\sidecaption
\includegraphics[width=0.75\textwidth,clip]{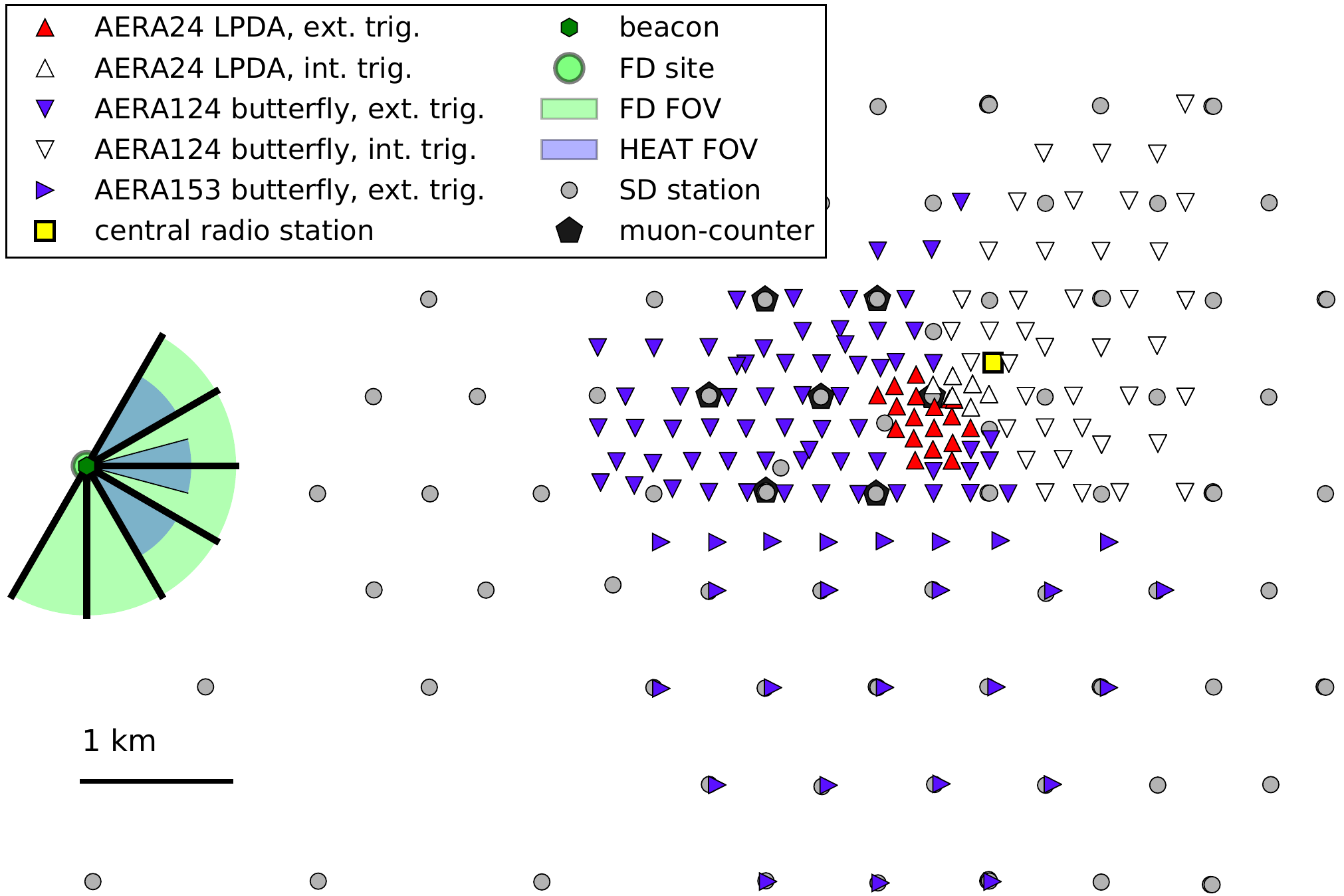}
\caption{Map of AERA within the Pierre Auger Observatory. The radio detector stations (triangles) equipped with different antennas and digitizing hardware are surrounded by surface detector (SD) stations (gray filled circles) and underground muon counters (black pentagons). AERA is in the field of view of the Coihueco and HEAT fluorescence telescopes. Fig. adapted from \cite{ICRC2015JSchulz}.}
\label{fig:AERAmap}       
\end{figure}

A map of the array is presented in Fig.~\ref{fig:AERAmap}. The AERA stations are running with slightly different types of electronics. Most of the stations (filled triangles) are equipped with a ring buffer large enough to store the measured radio signals for \SI{7}{s} to be able to receive an external trigger from the SD and FD. The rest of the stations which are equipped with a Butterfly antenna contain small scintillators for trigger purposes.

\section{Scientific Potential}
The different antenna and electronics designs implemented in AERA and the location of AERA next to additional detectors of the Pierre Auger Observatory where air showers are measured with three independent detection techniques create a large engineering and scientific potential. 

AERA can be used to explore the optimal setup to measure cosmic rays. Two different antenna designs are tested in the harsh environment of the Argentinian Pampa. The different detector spacings allow the determination of the optimal layout for different energy regimes, e.g., the \SI{750}{m} spaced stations are located directly next to SD stations to test the potential of a large-scale extension of a surface particle detector with radio antennas. The different types of station hardware allow the implementation of different trigger schemes such as self-triggering directly on the radio signal using advanced signal processing in an FPGA, external triggering or an internal particle trigger using the small scintillators integrated into the station design.  The latter setup is especially interesting for a possible future stand-alone radio array. 

The Auger surface detector, which is calibrated with the fluorescence detector, provides an accurate measurement of the cosmic-ray energy and is used to develop and cross-calibrate the energy measurement with AERA (see Sec.~\ref{sec:energycalib} or \cite{ICRC2015CGlaser, AERAEnergyPRL, AERAEnergyPRD}). Furthermore, the position of the shower maximum $X_\mathrm{max}$, which is an estimator of the primary-particle type of the cosmic ray, is directly measured with the fluorescence telescopes. Coincident measurements of AERA with FD are thus used to evaluate the performance of the different techniques used at AERA to reconstruct the cosmic-ray particle type. Different methods are developed at present to derive the cosmic-ray coposition from the radio measurements \cite{ARENA2016Gate}. 

In addition to a reconstruction of $X_\mathrm{max}$ using solely radio data, the AERA measurement can be combined with the other detectors. The radio signal measures the electromagnetic component of the air shower. Hence, a coincident measurement of the muonic shower component using underground particle detectors is sensitive to the cosmic-ray particle type. In case of horizontal air showers, the electromagnetic part of the air shower is completely absorbed in the atmosphere so that the SD measures only the muonic shower component whereas AERA still measures the electromagnetic part of the shower as the atmosphere is transparent to radio waves. Hence, a combined radio and particle measurement of horizontal air showers is sensitive to the cosmic-ray particle type \cite{ARENA2016Kambeitz}.

\section{In-situ Calibration of Antenna Response}
The most challenging part of the detector calibration is the absolute calibration of the antenna response to cosmic-ray radio signals. To be unaffected by near-field effects, a calibrated signal source needs to be placed at $\SI{\sim 20}{m}$ distance at any place around the antenna. Several calibration campaigns were performed and the measurement setup was continuously improved. The first campaign used a weather balloon to lift the calibrated signal source and measured the antenna response with an accuracy of 12.5\% \cite{AntennaPaper, AERAEnergyPRD}. The disadvantages of this method are the high costs of each balloon and the need for a windless environment. Therefore a new method that uses a remotely piloted drone to lift the signal source was developed. This setup was completed with an optical camera system to accurately determine the position of the signal source during the flight. Using this new method an accuracy of 9.3\% was achieved for the LPDA antenna \cite{Krause2016, ARENA2016Briechle}.

\section{Measurement of the Radiation Energy of Extensive Air Showers}
\label{sec:energycalib}
\begin{figure}[tb]
\centering
\includegraphics[width=0.55\textwidth,clip]{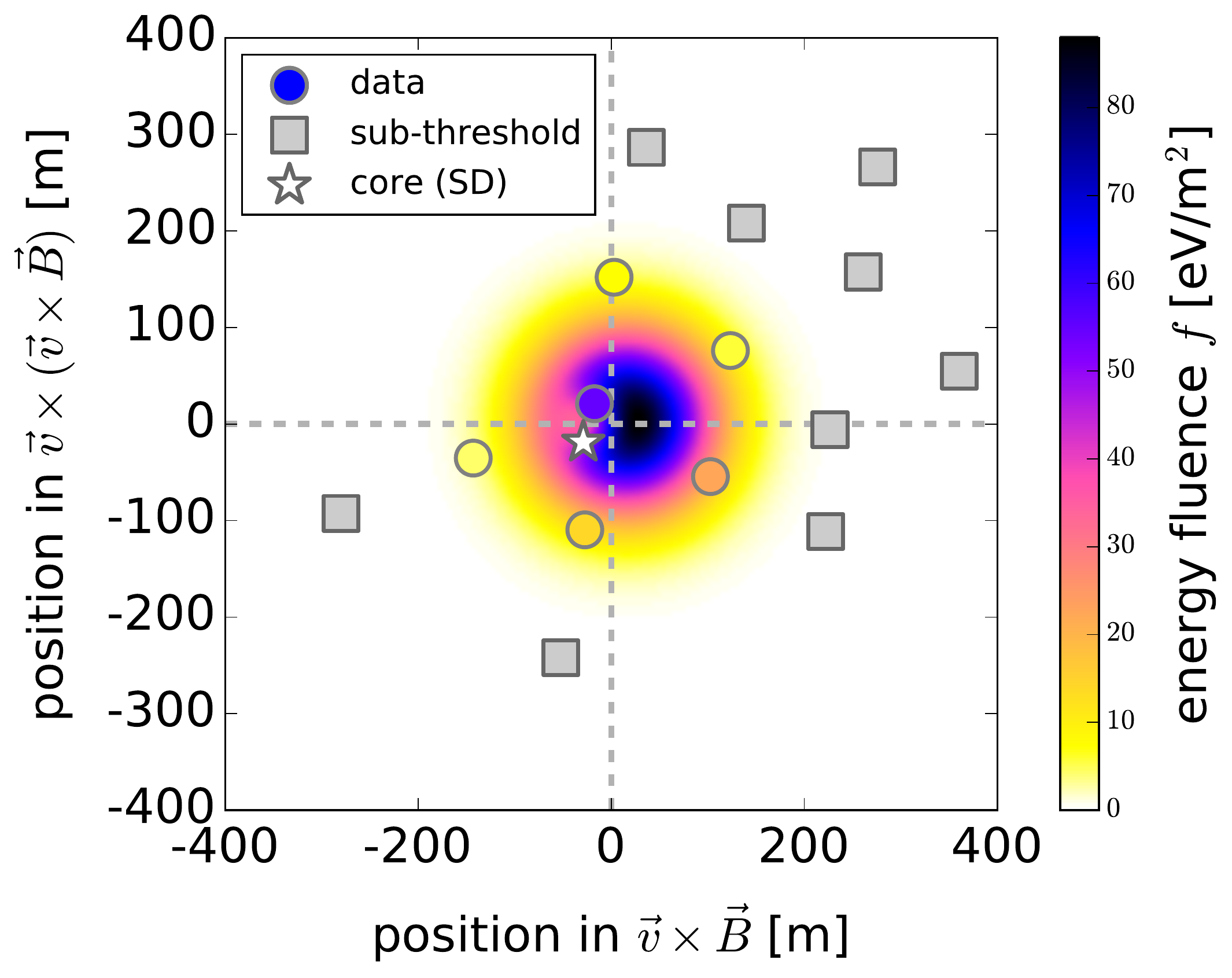} \hspace{0.1cm}
\includegraphics[width=0.43\textwidth,clip]{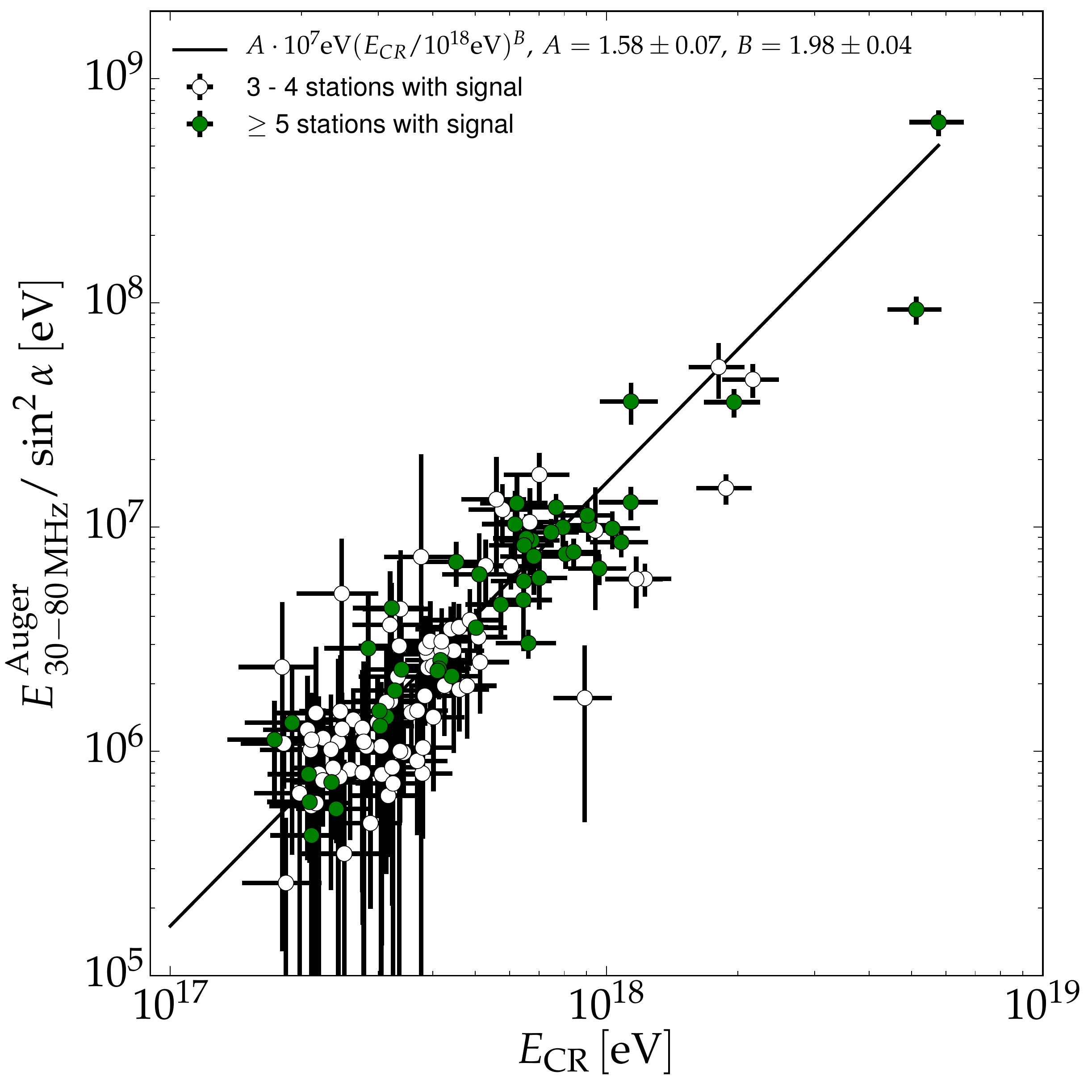}
\caption{(left) Energy fluence for an extensive air shower with
an energy of \SI{4.4E17}{eV} and a zenith angle of 25$^\circ$ as
measured in individual AERA radio detectors (circles filled
with color corresponding to the measured value) and fitted
with the azimuthally asymmetric, two-dimensional signal distribution function (background color).
The fit is performed in the plane perpendicular to the shower axis, with the $x$ axis
oriented along the direction of the Lorentz force for charged
particles propagating along the shower axis $\mathbf{v}$ in the geomagnetic field $\mathbf{B}$. The best-fitting impact point of the air shower
is at the origin of the plot, slightly offset from the one
reconstructed with the Auger surface detector [core (SD)].
(right) Correlation between the normalized radiation energy
and the cosmic-ray energy $E_\mathrm{CR}$ as determined by the Auger
surface detector. Open circles represent air showers with radio
signals detected in three or four radio detectors. Solid circles
denote showers with five or more detected radio signals. Figures and Caption from \cite{AERAEnergyPRL}.
}
\label{fig:LDF}       
\end{figure}

To measure the radiation energy, i.e., the amount of energy that is transferred from the cosmic ray during the air shower development into radio emission in the frequency band of AERA (\unit[30 - 80]{MHz}), we first unfold the detector response from the measured voltage traces. Thereby we obtain the electric field of the cosmic-ray radio signal from which we calculate the energy fluence, i.e., the energy in the radio signal per unit area, at each radio station. A two-dimensional function \cite{LOFARLDF, AERAEnergyPRD} is fitted to the signal distribution and integrated to obtain the radiation energy of the air shower (cf. Fig.~\ref{fig:LDF} left).

We correct the radiation energy for the different emission strengths at different geomagnetic angles $\alpha$ by dividing the radiation energy by $\sin^2 \alpha$. This corrected radiation energy correlates with the energy of the cosmic ray. Comparing the radiation energy with a measurement of the cosmic-ray energy by the surface detector we found that the radiation energy scales quadratically with the cosmic-ray energy, and we measured a radiation energy of \unit[$15.8 \pm 0.7 \mathrm{(stat)} \pm 6.7 \mathrm{(sys)}$]{MeV} for a \SI{1}{EeV} air shower arriving perpendicular to the geomagnetic field (cf. Fig.~\ref{fig:LDF} right). 

We can generalize our measurement to be applicable at any place on Earth by normalizing the radiation energy to the geomagnetic field strength. Then this corrected radiation energy is the same at any experiment and our result can be used by others to calibrate against the accurate energy scale of the Pierre Auger Observatory. More details can be found in \cite{ICRC2015CGlaser, AERAEnergyPRD, AERAEnergyPRL}. 

\section{Independent Determination of Cosmic-Ray Energy Scale from First Principles}
The measurement of the radiation energy can also be used for an independent determination of the cosmic-ray energy scale. The radiation energy released by air showers can be calculated from first principles. The radio emission originates only from the well-understood electromagnetic part of the air shower and is due to (de)acceleration of charges which is described by classical electrodynamics. Such a calculation is implemented in modern Monte-Carlo simulations \cite{ZHAires2012, CoREAS2013} which can be used for an accurate prediction of the radiation energy \cite{AERAEnergyPRL, GlaserErad2016, ARENA2016GlaserErad}.

Compared to a measurement of fluorescence light, the measurement of radiation energy has the advantage that it is hardly influenced by environmental conditions. The atmosphere is transparent to radio waves in the frequency range of AERA and also the amount of radiation energy released by an air showers is barely influenced by the current state of the atmosphere \cite{GlaserErad2016}.

\section{Conclusions}

The Auger Engineering Radio Array is the world's largest radio detector for high-energy cosmic rays consisting of 153 autonomously-operating radio stations covering an area of \SI{17}{km^2}. 
Its location within the Pierre Auger Observatory creates a large scientific potential as air showers are measured with four independent detection techniques simultaneously. The radiation energy of air showers was measured and compared with the accurate cosmic-ray energy information of the Auger surface detector. Furthermore, the radiation energy can be calculated from first principles for an independent determination of the cosmic-ray energy scale.

\end{document}